# Reversed Gravitational Acceleration for High-speed Particles

Hans C. Ohanian

Department of Physics

University of Vermont, Burlington, VT 05405-0125, USA

November 14, 2011

Abstract. Examination of the free-fall motion of particles of extremely high-speed in the Schwarzschild geometry reveals that the gravitational acceleration of such particles is reversed when measured in Schwarzschild coordinates. High-speed particles *decelerate* when moving radially downward, and they *accelerate* when moving upward. The onset of this abnormal behavior occurs at a speed of $1/\sqrt{3}$ times the local value of the speed of light. However, the gravitational force always remains attractive.

Einstein first discovered the gravitational time dilation in 1911, on the basis of his equivalence principle. From the time dilation, he immediately deduced the slow-down and deflection of light in a gravitational field.[1] His 1911 result for the reduction of the speed of light was in error by a factor of 2, but he corrected this a few years later, in his theory of General Relativity.

However, Einstein and his followers failed to notice a curious consequence of the slow-down of the speed of light: high-speed particles entering a gravitational field also slow down, and their gravitational acceleration is reversed, that is, the particles *decelerate when moving radially downward* in the gravitational field of a massive body, and they *accelerate when moving upward*.[2] It is easy to understand why this must happen in the case of an ultrarelativistic particle of initial speed $v \cong 1$. When this particle descends in the gravitational field of a massive body, it must obey the speed limit set by the decreasing speed of light, and this compels it to decelerate in the same way as light. Of course, a low-speed particle entering this gravitational field will accelerate in the normal way. It is then obvious that there is some critical speed $v_{crit} < 1$ that serves as a criterion for acceleration vs. deceleration: particles of speed smaller than $v_{crit}$ accelerate, but particles of speed larger than $v_{crit}$ decelerate.

---

[1] A. Einstein, Annalen d. Physik **35**, 898 (1911).

[2] The case of infall of high-speed particles seems not to have been treated in the available literature. C. W. Misner, K. S. Thorne, and J. A. Wheeler, *Gravitation* (Freeman, San Francisco, 1973 ) treat radial infall in detail, but only for the case of particles that start from rest at a given initial radius *R* (pp. 663- 668).

Here I will show that, in the Schwarzschild geometry, the critical speed for reversal of the radial acceleration is $v_{crit} = 1/\sqrt{3}$ times the local, slowed speed of light. This relationship is independent of position and independent of the value of the mass.

In this context, the decelerating speed is, of course, a coordinate speed, as is the slowed speed of light. That is, the speed is the ratio $dr/dt$ of the changes in Schwarzschild radial and time coordinates. But this coordinate speed is not devoid of physical significance. The coordinate speed $dr/dt$ can be readily determined by means of measurements with instruments placed in the far-away region, for instance, by radar-ranging, with the instantaneous distance calculated according to the time-delay and the known formula for the coordinate speed of light. For a radar pulse traveling radially, emitted at $r_0$ and reflected by the particle at $r$, the round trip travel time $\Delta t$ is given by[3]

$$\Delta t = 2 \int_{r_0}^{r} \frac{dr'}{1 - 2GM/r'}$$

or

$$\Delta t = 2(r_0 - r) + 4GM \ln\left(\frac{r_0/2GM - 1}{r/2GM - 1}\right) \quad (1)$$

from which the coordinate $r$ can be evaluated immediately. Thus, this coordinate $r$ has the status of a measurable quantity, and so does the speed $dr/dt$ calculated from its rate of change.

For a particle falling radially, the equation of geodesic motion in the Schwarzschild geometry[4] reduces to a simple expression, reminiscent of the Newtonian equation:

$$\frac{d^2 r}{d\tau^2} = -\frac{GM}{r^2} \quad (2)$$

Accordingly, for radial infall, the proper speed $|dr/d\tau|$ always increases. But the coordinate speed $|dr/dt|$ differs from this proper speed by a factor $d\tau/dt$ which is smaller than 1 and decreasing. For a high-speed particle, the decrease of this factor can overwhelm the increase of $|dr/d\tau|$ and lead to a decrease of $|dr/dt|$, that is, a deceleration.

---

[3] W. Rindler, *Essential Relativity* (Springer-Verlag, New York, 1977), p. 142.
[4] See, e.g., H. C. Ohanian and R. Ruffini, *Gravitation and Spacetime* (W. W. Norton & Co, New York, 1994), p. 401.

The left side the equation of motion is

$$\frac{d^2r}{d\tau^2} = \frac{d}{d\tau}\left(\frac{dt}{d\tau}\frac{dr}{dt}\right) = \frac{dt}{d\tau}\frac{d^2r}{d\tau dt} + \frac{dr}{dt}\frac{d}{d\tau}\left(\frac{dt}{d\tau}\right)$$

With the substitution

$$\frac{dt}{d\tau} = \frac{1}{\sqrt{(1-2GM/r)-(dr/dt)^2/(1-2GM/r)}} \quad (3)$$

this becomes

$$\frac{d^2r}{d\tau^2} = \frac{1}{(1-2GM/r)^2 - v^2}\left[\left(1-\frac{GM}{r}\right)^2 \frac{d^2r}{dt^2} - \frac{GM}{r^2}v^2 - \frac{GM}{r^2}\frac{1}{(1-2GM/r)^2}v^4\right]$$

where $v \equiv dr/dt$. Accordingly, the equation of motion becomes

$$\frac{1}{(1-2GM/r)^2 - v^2}\left[\left(1-\frac{GM}{r}\right)^2 \frac{d^2r}{dt^2} - \frac{GM}{r^2}v^2 - \frac{GM}{r^2}\frac{1}{(1-2GM/r)^2}v^4\right] = -\frac{GM}{r^2}$$

The critical speed $v_{crit}$ is determined by the condition $d^2r/dt^2 = 0$ which implies

$$\left[-\frac{GM}{r^2}v^2 - \frac{GM}{r^2}\frac{1}{(1-2GM/r)^2}v^4\right] = -\frac{GM}{r^2}\left[(1-2GM/r)^2 - v^2\right] \quad (4)$$

This is a quadratic equation for $v^2$, with the solution

$$v_{crit} = \frac{1}{\sqrt{3}}\left(1-\frac{2GM}{r}\right) \quad (5)$$

Since $(1-2GM/r)$ is the local, slowed speed of light, this says that the critical speed is $1/\sqrt{3}$ times the local speed of light.

Figure 1 shows plots of the speed $v = |dr/dt|$ of freely-falling particles as a function of the radial coordinate. [The speed is determined by the first integral of the equation of motion (2),

$$\frac{1}{2}\left(\frac{dr}{d\tau}\right)^2 = \frac{GM}{r} + const. = \frac{GM}{r} + \frac{1}{2}\frac{v_0^2}{1-v_0^2} \qquad (6)$$

where $v_0$ is the initial speed of the particle at large distance. The combination of Eqs. (6) and (3) then gives the value of $dr/dt$,

$$\left(\frac{dr}{dt}\right)^2 = \frac{[2GM/r + v_0^2/(1-v_0^2)](1-2GM/r)}{1+[2GM/r + v_0^2/(1-v_0^2)]/(1-2GM/r)} \quad ] \qquad (7)$$

Note that for an initial speed $v_0 = 1/\sqrt{3}$ (not plotted in Fig. 1), the particle proceeds with constant speed as long as the linear approximation for the gravitational field is valid, that is, for $GM/r \ll 1$ (within this linear regime, the critical speed is simply $v_{crit} \cong 1/\sqrt{3}$ times the standard speed of light). The particle then decelerates when it enters the nonlinear regime of the Schwarzschild geometry.

Also note that for a particle moving in a transverse, or tangential, direction, the radial acceleration is always downward, that is, the acceleration does not reverse at high speed. Thus, such a particle deflects in the normal way and, for a particle of speed $v \cong 1$, the deflection is the same as for a light signal.

Taken at face value, the discrepancy between the signs of the accelerations of low-speed and high-speed particles is a perplexing violation of the equivalence principle. General Relativity attributes this discrepancy to a bad choice of coordinates—the coordinates $r$ and $t$ do not represent locally measured distances and times. In local geodesic coordinates, with $\Gamma^\mu_{\alpha\beta} = 0$, the accelerations of all particles are zero, and the discrepancy disappears.

In 1911, Einstein would not have known about this way of avoiding the violation of the equivalence principle. If he had noticed that the slowed speed of light requires a slowed speed for ultrarelativisic particles, he would have been in a quandary. But he didn't notice, and neither did anybody else (until several years later; see Correction attached at end of this paper).

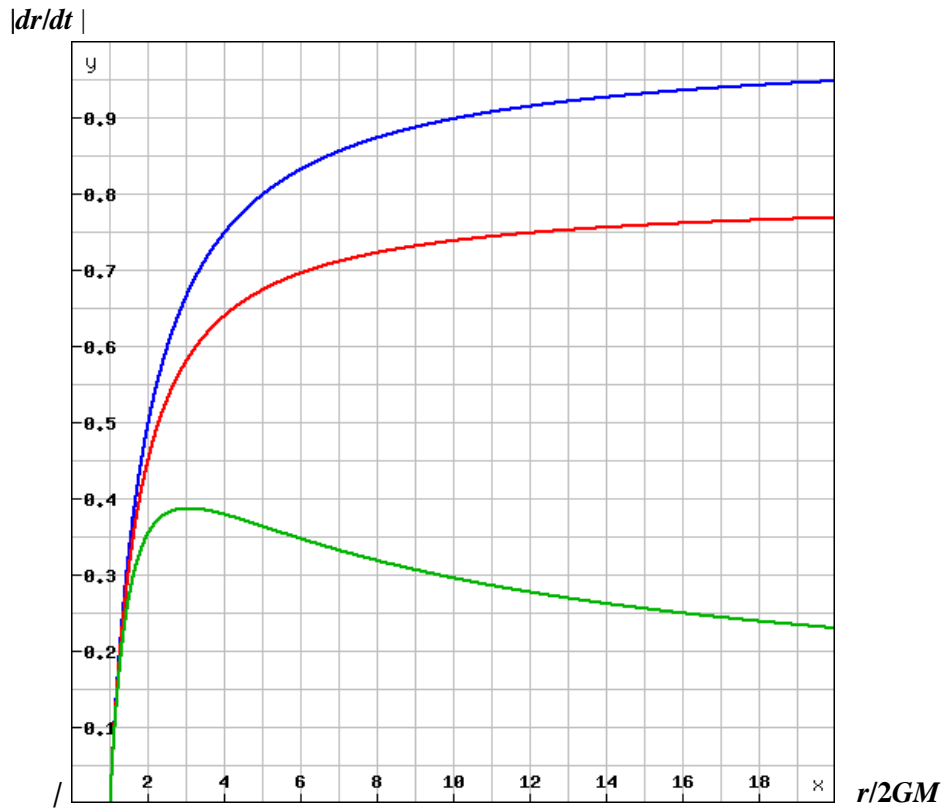

**Fig. 1** Speed of a freely-falling particle vs. radial coordinate in units of 2*GM*. *Upper curve*: light signal. *Middle curve*: particle of initial speed more than $v_{crit}$ ; the particle decelerates monotonically as it falls downward from a large initial distance.  *Lower curve*: particle of initial speed less than $v_{crit}$ ; the particle accelerates until its speed reaches the critical value at a radial coordinate of $\cong 3.1$ units, and it then decelerates. The value of $v_{crit}$ at the peak is $1/\sqrt{3}$ times the local, slowed speed of light.

*Corrrection and Addendum*

Contrary to my assertion that Einstein's followers failed to notice the deceleration of high-speed particles entering a gravitational field, there are actually a handful of publications that discuss this deceleration, ranging from a 1917 paper by Hilbert[5] to recent papers by Mashoon[6] and by Felber[7] (McGruder[8] gives a comprehensive list of papers before 1982). I am indebted to B. Mashoon, F. Felber, and J. Moore for bringing these publications to my attention.

Hilbert, a better mathematician than physicist, unfortunately misconstrued this deceleration as a repulsive gravitational force ("die Gravitation wirkt…abstossend"), and some authors imitated this mistake. Hilbert naively assumed that the force is in the direction of the coordinate acceleration $d^2r/dt^2$, whereas he should have known that the force is in the direction of the rate of change of momentum. According to Eq. (2), the rate of change of the relativistic momentum is

$$\frac{d}{d\tau} p_r = \frac{d}{d\tau}\left( m \frac{dr}{d\tau} \right) = -\frac{GMm}{r^2} \tag{8}$$

which is always negative. Therefore the direction of the force (and also the direction of the proper, or relativistic, acceleration $d^2r/d\tau^2$) is always downward, that is, *the force is always attractive*. As already mentioned on p. 2, for a high-speed particle moving downward, $d^2r/dt^2$ and $d^2r/d\tau^2$ can have opposite signs, because $|dr/dt|$ differs from $|dr/d\tau|$ by a factor $d\tau/dt$, and the decrease of this factor can overwhelm the increase of $|dr/d\tau|$ and lead to a decrease of $|dr/dt|$, that is, a deceleration. But the sign of $d^2r/dt^2$ does not determine the sign of the force, which always remains attractive and increases in magnitude with decreasing *r*, according to the inverse-square law (8).

Hilbert's repulsive force is a delusion that rests on bad physics, and Felber's contention that this repulsive force can be exploited for an "antigravity" spacecraft propulsion scheme rests on equally bad physics. Felber's scheme[7] is merely a relativistic version of the familiar Newtonian "slingshot" effect that has been used to boost the terminal speed (and momentum) of several spacecraft by using orbits that swing the spacecraft around a moving planet; this involves the gravitational attraction of the planet, not any kind of repulsion.

---

[5] D. Hilbert, Nachrichten Königl. Ges. Wiss. Göttingen, 1917, p. 53, available at http://resolver.sub.uni-goettingen.de/purl?GDZPPN002504561 . The reversal of acceleration is mentioned in passing on the last page of this paper.
[6] B. Mashoon, Int. J. Mod. Phys. D **14**, 2025 (2005).
[7] F. Felber, AIP Conference Proceedings **1208**, 247 (2010), also available at arXiv:0910.1084v2.
[8] C. H. McGruber, III, Phys. Rev. D **25**, 3191 (1982).